# Statistical Thermodynamics and Ordering Kinetics of $D0_{19}$-Type Phase: Application of the Models for H.C.P.-Ti–Al Alloy

T.M. Radchenko[1,a], V.A. Tatarenko[1,b] and H. Zapolsky[2,c]

[1]Department of Solid State Theory, Institute for Metal Physics, N.A.S. of Ukraine,
36 Academician Vernadsky Boulevard, 03680 Kyiv-142, Ukraine

[2]Institut des Matériaux, UFR Sciences et Techniques, Université de Rouen,
Avenue de l'université–B.P. 12, 76801 Saint-Etienne-du Rouvray Cedex, France

[a]Taras.Radchenko@gmail.com, [b]tatar@imp.kiev.ua, [c]Helena.Zapolsky@univ-rouen.fr

**Keywords:** Order–disorder transformation, ordering kinetics, interatomic interaction.

**Abstract.** Using the self-consistent field approximation, the static concentration waves approach and the Onsager-type kinetics equations, the descriptions of both the statistical thermodynamics and the kinetics of an atomic ordering of $D0_{19}$ phase are developed and applied for h.c.p.-Ti–Al alloy. The model of order–disorder phase transformation describes the phase transformation of h.c.p. solid solution into the $D0_{19}$ phase. Interatomic-interaction parameters are estimated for both approximations: one supposes temperature-independent interatomic-interaction parameters, while the other one includes the temperature dependence of interchange energies for Ti–Al alloy. The partial Ti–Al phase diagrams (equilibrium compositions of the coexistent ordered $\alpha_2$-phase and disordered $\alpha$-phase) are evaluated for both cases. The equation for the time dependence of $D0_{19}$-type long-range order (LRO) parameter is analyzed. The curves (showing the LRO parameter evolution) are obtained numerically for both temperature-independent interaction energies and temperature-dependent ones. Temperature dependence of the interatomic-interaction energies accelerates the LRO relaxation and diminishes a spread of the values of instantaneous and equilibrium LRO parameters versus the temperature. Both statistical-thermodynamics and kinetics results show that equilibrium LRO parameter for a non-stoichiometry (where an atomic fraction of alloying component is more than 0.25) can be higher than for a stoichiometry at high temperatures. The experimental phase diagram confirms the predicted (ordered or disordered) states for h.c.p.-Ti–Al.

## Introduction

Ordering of solid solutions is a topic of growing technological and scientific interest within the field of materials science. This is because of the advantageous high-temperature and corrosion properties of alloys (in particular, Ti–Al intermetallic compound) are linked to the effect of long-range ordering [1]. The LRO kinetics is one of the microdiffusion processes occurring within the atomic ranges. X-ray diffraction technique is the most convenient experimental instrument to provide us detailed information about this process. However, such experimental measurements have not been done for some high-temperature hexagonal closed-packed (h.c.p.) alloys, for instant, for Ti–Al intermetallics. Therefore, as yet, theoretical investigation is a better way.

Intermetallic compounds based on alloys of the light elements, Ti and Al, are promising candidates for applications as high-temperature engineering materials due to some properties such as low density, high melting point, high strength and creep resistance, high oxidation resistance, low diffusion coefficients etc. $Ti_3Al$ is one of the intermetallic compounds, which have received much attention as advanced materials for applications as high-temperature structural materials in aerospace and automobile industries [2–5]. Ti–Al alloys used in aircraft construction are three times stronger than steel and 45 percent lighter. The $Ti_3Al$ phase is basic during the creation of industrial titanium materials and provides a beneficial effect for the mechanical properties of Ti–Al based alloys.

The structure of $\alpha_2$-Ti$_3$Al phase is related to the D0$_{19}$-type superstructure [6–16], which can be derived from the h.c.p. solid solution ($\alpha$-Ti–Al) by the ordering of substitutional Ti and Al atoms. The $\alpha_2$-Ti$_3$Al phase is an ordered structure, has been an object of many-years' discussions concerning its character and formation. Different variants of Ti–Al phase diagram have been proposed too.

The first step in an understanding of microstructure and microdiffusion kinetics of the Ti–Al alloy relaxation is the construction of a statistical-thermodynamic model and the estimation of interatomic-interaction energy parameters for this system. In spite of some efforts [17], which have been made to describe the atomic-ordering reaction in Ti$_3$Al ($\alpha \rightarrow \alpha_2$), the statistical-thermodynamic description for non-stoichiometric h.c.p. phase has not been attempted. Description of the site occupation of alloying element in the ordered intermetallic compounds is essential for further understanding of alloy behavior such as the toughening, strengthening and stabilizing effect of alloying elements. However, only a few experimental results and theoretical descriptions have been obtained [10,18–20] to describe the site occupation in Ti–Al alloys. Atomic-distribution functions for substitutional superstructures in h.c.p. lattices have been obtained in Ref. [21], however, LRO parameters have not been calculated. The spatial-distribution functions for atoms within the D0$_{19}$ superstructure given in Refs. [7,22,23] as well as the image of D0$_{19}$ superstructure presented in [24,25] are wrong, in spite of the fact that co-ordinates of the atoms are correctly indicated in [24,25].

A given paper is concerned, firstly, with the statistical-thermodynamic description of phase transformation of the disordered h.c.p. solid solution into the ordered D0$_{19}$-type phase and, secondly, with the relaxation kinetics of the LRO parameter during the D0$_{19}$-phase ordering. The Ti–Al alloy is chosen as a case in point. The statistical-thermodynamics model is based on the self-consistent field approximation and the static concentration wave approach [26]. The kinetics model is based on the Onsager-type kinetic equation [26–34]. Kinetics results in Refs. [26–34] relate to the alloys based on cubic Bravais lattice. Therefore, investigation of ordering kinetics for non-cubic complicated Ising lattice is relevant to present time.

The paper consists of a few sections and subsections. The statistical-thermodynamics model of binary substitutional h.c.p. solid solution is presented in the next section. At first, consider the general case of h.c.p.-crystal lattice and then the D0$_{19}$-type structure in Ti–Al have been considered. Then the LRO-kinetics model for D0$_{19}$-phase was analyzed. The subsequent section is concerned with the statistical-thermodynamics and kinetics results along with the corresponding conclusions. Finally, the last section contains the summary.

**Statistical-Thermodynamics Model of Binary H.C.P. Substitutional Alloy**

**General Case.** The hexagonal close-packed lattice is a complicated Ising lattice, which can be considered as two interpenetrating hexagonal Bravais sublattices displaced with respect to each other by the vector $\mathbf{h} = 2\mathbf{a}_1/3 + \mathbf{a}_2/3 + \mathbf{a}_3/2$, where $\mathbf{a}_1, \mathbf{a}_2, \mathbf{a}_3$ are the primitive-translation vectors of h.c.p. lattice along the [100], [010], [001] directions, respectively, in the oblique system of coordinates (see Fig. 1). Two vectors $\mathbf{R}$ and $\mathbf{h}_p$ can characterize an each crystal lattice site $\mathbf{r}$: $\mathbf{R} + \mathbf{h}_p = \mathbf{r}$ [26]. Vector $\mathbf{R}$ refers to the origin-site position within the primitive unit cell (where a given site $\mathbf{r}$ is located); $\mathbf{h}_p$ is the radius-vector of a given site with respect to the unit-cell's origin; an index $p = 1, 2$ denotes the sublattices. It is easy to see from Fig. 1 that $\mathbf{h}_1 = \mathbf{0}$ and $\mathbf{h}_2 = 2\mathbf{a}_1/3 + \mathbf{a}_2/3 + \mathbf{a}_3/2$.

Within the scope of the self-consistent field approximation, the configurational part of free energy of the binary h.c.p.-$A$–$B$ alloy based on the complicated Ising lattice can be written [26] as

$$F = \frac{1}{2}\sum_{p,q=1}^{2}\sum_{\mathbf{R},\mathbf{R}'} w_{pq}(\mathbf{R}-\mathbf{R}')P_p(\mathbf{R})P_q(\mathbf{R}') + k_B T \sum_{q=1}^{2}\sum_{\mathbf{R}}[P_q(\mathbf{R})\ln P_q(\mathbf{R}) + (1-P_q(\mathbf{R}))\ln(1-P_q(\mathbf{R}))], \quad (1)$$

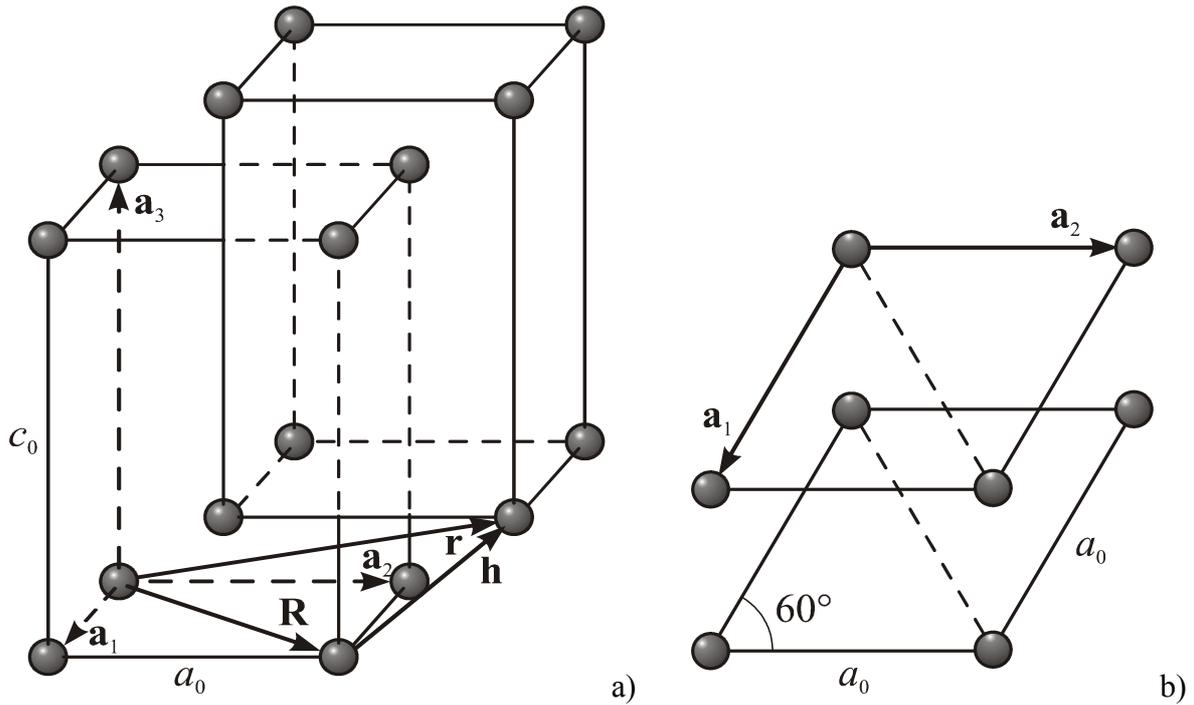

Fig. 1 Hexagonal close-packed lattice: perspective view (a) and top view (b).

where the indexes $p$ and $q$ denote the sublattices ($p, q = 1, 2$), $k_B$ is the Boltzmann constant, and $T$ is the absolute temperature of an alloy. The single-site occupation-probability function $P_p(\mathbf{R})$ ($P_q(\mathbf{R})$) represents the probability of finding a $B$ atom at the site of the $p$-th ($q$-th) sublattice within the cell with the origin at $\mathbf{R}$. In the last expression, the summation is carried out over the all primitive unit cells ($\mathbf{R}, \mathbf{R}'$) and all sublattices, *i.e.* over all Ising lattice sites.

For a binary solid solution, $w_{pq}(\mathbf{R} - \mathbf{R}')$ defines the interchange energy [26], which is also known as mixing energy:

$$w_{pq}(\mathbf{R} - \mathbf{R}') = W_{pq}^{AA}(\mathbf{R} - \mathbf{R}') + W_{pq}^{BB}(\mathbf{R} - \mathbf{R}') - 2W_{pq}^{AB}(\mathbf{R} - \mathbf{R}'). \qquad (2)$$

Here, $W_{pq}^{AA}$, $W_{pq}^{BB}$ and $W_{pq}^{AB}$ are the pairwise-interaction energies of $A$–$A$, $B$–$B$ and $A$–$B$ pairs of atoms, respectively, for the unit cells separated by a distance $|\mathbf{R} - \mathbf{R}'|$. The radius-vector $\mathbf{R}$ is related to the basic vectors as $\mathbf{R} = n_1 \mathbf{a}_1 + n_2 \mathbf{a}_2 + n_3 \mathbf{a}_3$ with $|\mathbf{a}_1| = |\mathbf{a}_2| = a_0$, $|\mathbf{a}_3| = c_0$; $n_1$, $n_2$, $n_3$ are the integer 'co-ordinates' of the unit-cell origins within the oblique co-ordinates of h.c.p. lattice (Fig. 1).

To determine the spatial-distribution function, $P_q(\mathbf{R})$, the static concentration waves approach [26] was used. Within the scope of this approach, the distribution function can be represented as a linear superposition of the static concentration waves:

$$\varphi_{\sigma \mathbf{k}}(q, \mathbf{R}) = v_\sigma(q, \mathbf{k}) e^{i \mathbf{k} \cdot \mathbf{R}}, \qquad (3)$$

where $\mathbf{k}$ is a wave vector, $\|v_\sigma(q, \mathbf{k})\|$ is a unit 'polarization vector' of the σ-th concentration wave, and σ is a 'polarization number'. The function $P_q(\mathbf{R})$ can be written in a form of the Fourier series:

$$P_q(\mathbf{R}) = c + \frac{1}{2} \sum_s \sum_{\sigma=1}^{2} \eta_{s,\sigma} \sum_{j_s} \left[ \gamma_{s,\sigma}(j_s) v_\sigma(q, \mathbf{k}_{j_s}) e^{i \mathbf{k}_{j_s} \cdot \mathbf{R}} + \gamma_{s,\sigma}^*(j_s) v_\sigma^*(q, \mathbf{k}_{j_s}) e^{-i \mathbf{k}_{j_s} \cdot \mathbf{R}} \right]. \qquad (4)$$

In the last expression, $c$ is an atomic fraction of $B$ atoms in $A_{1-c}B_c$ alloy; $v_\sigma(q, \mathbf{k}) e^{i \mathbf{k}_{j_s} \cdot \mathbf{R}}$ is a static concentration wave with the superlattice wave vector, $\mathbf{k}_{j_s}$, describing an ordered superstructure (index $j_s$ denotes the rays of the *s*-th wave-vector star within the first Brillouin zone of reciprocal space); $\{\eta_{s,\sigma}\}$ are the LRO parameters (they are equal to 0 or 1 in disordered state or completely-ordered one, respectively); $\gamma_{s,\sigma}(j_s)$ are coefficients determining a symmetry of the occupation probabilities, $P_q(\mathbf{R})$, i.e. the superstructure symmetry. The summation in expression (4) is carried out over all rays $\{j_s\}$ of the stars $\{\mathbf{k}^s\}$ and over all 'polarizations' $\{\sigma\}$.

The concentration waves, $\|\varphi_{\sigma\mathbf{k}}(q,\mathbf{R}')\|$, are eigenfunctions of the mixing-energy matrix, $\|w_{pq}(\mathbf{R}-\mathbf{R}')\|$:

$$\sum_{q=1}^{2}\sum_{\mathbf{R}'} w_{pq}(\mathbf{R}-\mathbf{R}')\varphi_{\sigma\mathbf{k}}(q,\mathbf{R}') = \lambda_\sigma(\mathbf{k})\varphi_{\sigma\mathbf{k}}(p,\mathbf{R}), \qquad (5)$$

where $\lambda_\sigma(\mathbf{k})$ is an eigenvalue of the matrix $\|w_{pq}(\mathbf{R}-\mathbf{R}')\|$. The 'polarization number' $\sigma$ is called as a number of branch in a spectrum, $\{\lambda_\sigma(\mathbf{k})\}$. Substitution of expression (3) into Eq. 5 gives the secular equation:

$$\sum_{p=1}^{2}\tilde{w}_{pq}(\mathbf{k})v_\sigma(p,\mathbf{k}) = \lambda_\sigma(\mathbf{k})v_\sigma(q,\mathbf{k}) \quad (q=1,2), \qquad (6)$$

where

$$\tilde{w}_{pq}(\mathbf{k}) = \sum_{\mathbf{R}} w_{pq}(\mathbf{R}-\mathbf{R}')e^{-i\mathbf{k}\cdot(\mathbf{R}-\mathbf{R}')} \qquad (7)$$

is the Fourier transform of a mixing energies. Since the Fourier-component matrix, $\tilde{w}_{pq}(\mathbf{k})$, is Hermitian, all its eigenvalues are real values, and eigenvectors, $\|v_\sigma(p,\mathbf{k})\|$, are orthogonal to each other,

$$\sum_{q=1}^{2} v_\sigma^*(q,\mathbf{k})v_{\sigma'}(q,\mathbf{k}) = \delta_{\sigma\sigma'}.$$

In expanded form, the $\tilde{w}_{pq}(\mathbf{k})$ matrix has a form [6,7,21,26] as follows:

$$\|\tilde{w}_{pq}(\mathbf{k})\| = \begin{pmatrix} \tilde{w}_{11}(\mathbf{k}) & \tilde{w}_{12}(\mathbf{k}) \\ \tilde{w}_{12}^*(\mathbf{k}) & \tilde{w}_{11}(\mathbf{k}) \end{pmatrix}, \qquad (8)$$

where $\tilde{w}_{12}^*(\mathbf{k})$ is complex conjugate to $\tilde{w}_{12}(\mathbf{k})$. The equalities $\tilde{w}_{11}(\mathbf{k}) = \tilde{w}_{22}(\mathbf{k})$ and $\tilde{w}_{21}(\mathbf{k}) = \tilde{w}_{12}^*(\mathbf{k})$, which are valid in case of the substitutional solutions based on the h.c.p. lattice [7,21–23] was used.

The following eigenvalues, $\lambda_\sigma(\mathbf{k})$, and eigenvectors, $\mathbf{v}_\sigma(\mathbf{k})$, correspond to the mixing-energy matrix $\|\tilde{w}_{pq}(\mathbf{k})\|$:

$$\lambda_1(\mathbf{k}) = \tilde{w}_{11}(\mathbf{k}) + |\tilde{w}_{12}(\mathbf{k})|, \qquad \lambda_2(\mathbf{k}) = \tilde{w}_{11}(\mathbf{k}) - |\tilde{w}_{12}(\mathbf{k})|, \qquad (9a)$$

$$\mathbf{v}_1(\mathbf{k}) = \frac{1}{\sqrt{2}}\begin{pmatrix} 1 \\ \dfrac{\tilde{w}_{12}^*(\mathbf{k})}{|\tilde{w}_{12}(\mathbf{k})|} \end{pmatrix}, \qquad \mathbf{v}_2(\mathbf{k}) = \frac{1}{\sqrt{2}}\begin{pmatrix} 1 \\ -\dfrac{\tilde{w}_{12}^*(\mathbf{k})}{|\tilde{w}_{12}(\mathbf{k})|} \end{pmatrix}. \qquad (9b)$$

The absolute instability (critical) temperature for the complicated Ising lattice is equal to [21,26]

$$T_c = -\frac{1}{k_B} c(1-c) \min_{\sigma,s} \lambda_\sigma(\mathbf{k}^s), \qquad (10)$$

where $\min_{\sigma,s} \lambda_\sigma(\mathbf{k}^s)$ is the absolute minimum of $\lambda_\sigma(\mathbf{k}^s)$. The phase transition is generated by the $\mathbf{k}^s$-star whose rays, $\{\mathbf{k}_{j_s}\}$, and 'polarization vectors', $\|v_\sigma(q,\mathbf{k})\|$, provide such a minimum of $\lambda_\sigma(\mathbf{k}_{j_s})$.

**Description of D0$_{19}$-Type Superstructure of Ti–Al Alloy.** Let us consider an ordered Ti$_3$Al lattice, which can be considered as D0$_{19}$-type superstructure—ordered distribution of Ti and Al atoms at the h.c.p.-lattice sites [6–16]. A unit cell of the ordered D0$_{19}$-Ti$_3$Al alloy consists of eight atoms at the following positions (see Fig. 2):

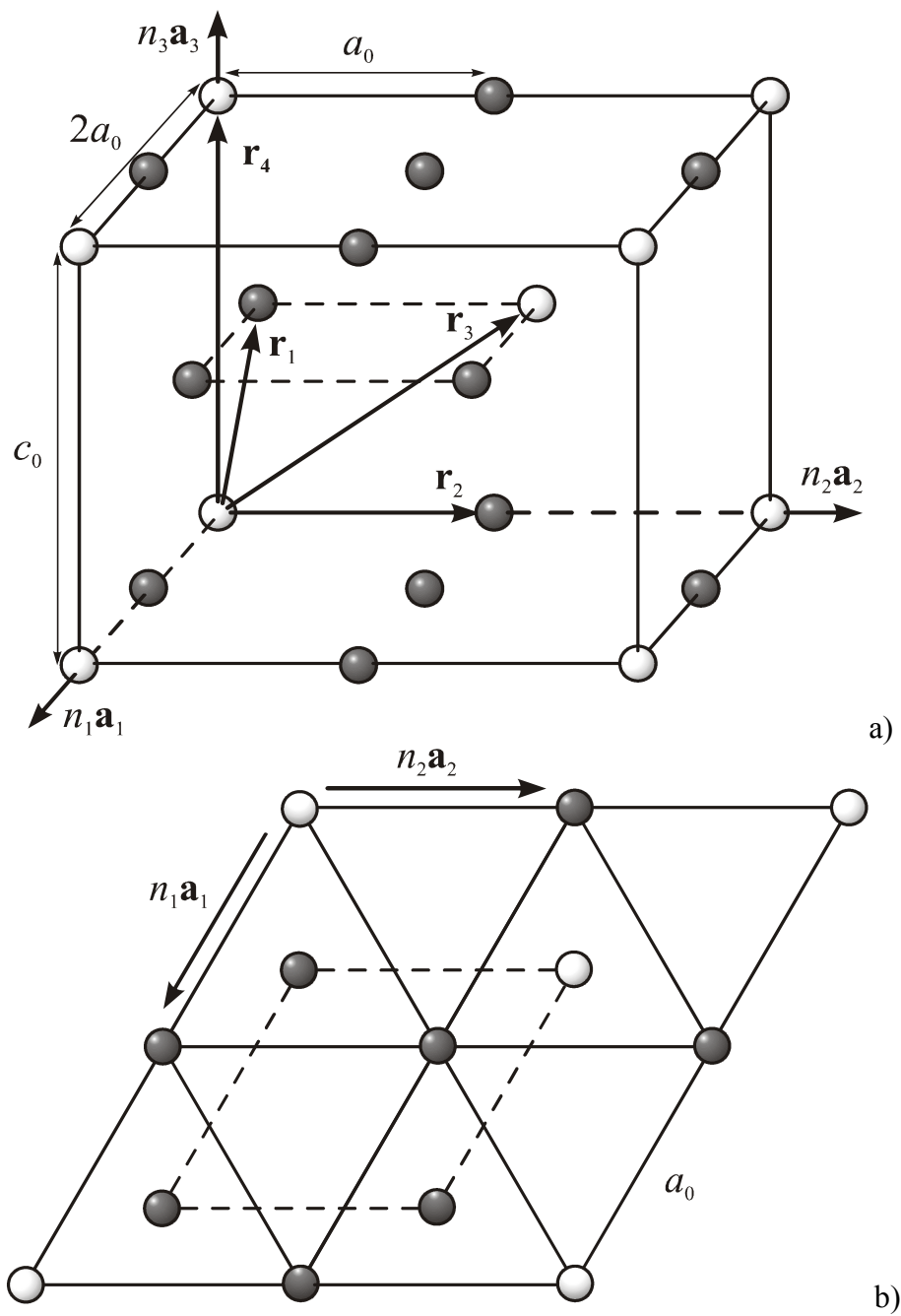

Fig. 2 A unit cell of D0$_{19}$-type Ti$_3$Al superstructure: perspective view (a) and top view (b). White balls correspond to Al atoms, dark balls—Ti atoms.

$$\begin{cases} \text{Al}: & (0\ 0\ 0),\ (2/3\ 4/3\ 1/2); \\ \text{Ti}: & (0\ 1\ 0),\ (1\ 0\ 0),\ (1\ 1\ 0); \\ \text{Ti}: & (2/3\ 1/3\ 1/2),\ (5/3\ 1/3\ 1/2),\ (5/3\ 4/3\ 1/2). \end{cases}$$

The lattice parameters, $a_0$ and $c_0$, for Ti$_3$Al were measured using the x-ray diffraction and confirmed by the electron diffraction [8]: $a_0 = 0.289$ nm, $c_0 = 0.464$ nm. If $\sqrt{6}c_0/4 < a_0 < 3c_0/4$, the radii of the first four co-ordination shells ($r_1$, $r_2$, $r_3$, $r_4$) are as follows (see Figs. 1, 2):

$$r_1 = \sqrt{a_0^2/3 + c_0^2/4},\qquad r_2 = a_0,\qquad r_3 = \sqrt{4a_0^2/3 + c_0^2/4},\qquad r_4 = c_0.$$

This means that $r_1 = 0.286$ nm, $r_2 = 0.289$ nm, $r_3 = 0.406$ nm and $r_4 = 0.464$ nm.

Here, it has to be noticed that some authors [13] assume that Ti$_3$Al lattice is a perfect h.c.p. lattice where $r_1 = r_2$ and $c_0/a_0 = 1.633$. Actually, $r_1 \neq r_2$. It is important for estimation of the mixing energies within both the first co-ordination shell and the second co-ordination shell.

Any h.c.p.-lattice reciprocal-space vector, $\mathbf{k}$, is $\mathbf{k} = (k_1, k_2, k_3) = 2\pi(k_1\mathbf{a}_1^* + k_2\mathbf{a}_2^* + k_3\mathbf{a}_3^*)$ with $\mathbf{a}_1^*$, $\mathbf{a}_2^*$ and $\mathbf{a}_3^*$ being the reciprocal-lattice translation vectors along the [100], [010] and [001] directions, respectively, and $|\mathbf{a}_1^*| = |\mathbf{a}_2^*| = 1/a_0$, $|\mathbf{a}_3^*| = 1/c_0$ for a h.c.p lattice. Using definition (7), one can obtain expressions for the elements of the matrix (8):

$$\tilde{w}_{11}(\mathbf{k}) = w_2[e^{2\pi i k_1} + e^{2\pi i k_2} + e^{-2\pi i k_1} + e^{-2\pi i k_2} + e^{2\pi i(k_1+k_2)} + e^{-2\pi i(k_1+k_2)}] + w_4[e^{2\pi i k_3} + e^{-2\pi i k_3}] + \ldots, \quad (11\text{a})$$

$$\tilde{w}_{12}(\mathbf{k}) = w_1[1 + e^{-2\pi i k_3} + e^{-2\pi i k_1} + e^{-2\pi i(k_1+k_3)} + e^{-2\pi i(k_1+k_2)} + e^{-2\pi i(k_1+k_2+k_3)}] +$$

$$+ w_3[e^{2\pi i k_2} + e^{2\pi i(k_2-k_3)} + e^{-2\pi i k_2} + e^{-2\pi i(k_2+k_3)} + e^{-2\pi i(2k_1+k_2)} + e^{-2\pi i(2k_1+k_2+k_3)}] + \ldots. \quad (11\text{b})$$

Here, $w_1$, $w_2$, $w_3$, $w_4$ are the mixing energies for the 1-st, 2-nd, 3-rd, 4-th co-ordination shells with the radii $r_1$, $r_2$, $r_3$, $r_4$ shown in Fig. 2, respectively.

The D0$_{19}$-type superstructure (Fig. 2) is generated by the rays $\{\mathbf{k}_{j_M}\}$ of superlattice wave-vector $\mathbf{k}^M$ [6,7,21,26,27]: $\mathbf{k}_1 = \pi\mathbf{a}_1^* = (1/2\ 0\ 0)$, $\mathbf{k}_2 = \pi\mathbf{a}_2^* = (0\ 1/2\ 0)$, $\mathbf{k}_3 = \pi(\mathbf{a}_1^* + \mathbf{a}_2^*) = (1/2\ 1/2\ 0)$. Using expressions (11a) and (11b), the elements of matrix (8) for these wave vectors as well as for zero wave vector $\mathbf{k}^\Gamma = \mathbf{0}$ can be written in the next forms:

$$\tilde{w}_{11}(\mathbf{0}) = 6w_2 + 2w_4 + \ldots,\qquad \tilde{w}_{12}(\mathbf{0}) = 6w_1 + 6w_3 + \ldots; \qquad (12\text{a})$$

$$\tilde{w}_{11}(\mathbf{k}_1) = -2w_2 + 2w_4 + \ldots,\qquad \tilde{w}_{12}(\mathbf{k}_1) = -2w_1 + 6w_3 + \ldots; \qquad (12\text{b})$$

$$\tilde{w}_{11}(\mathbf{k}_2) = -2w_2 + 2w_4 + \ldots,\qquad \tilde{w}_{12}(\mathbf{k}_2) = 2w_1 - 6w_3 + \ldots; \qquad (12\text{c})$$

$$\tilde{w}_{11}(\mathbf{k}_3) = -2w_2 + 2w_4 + \ldots,\qquad \tilde{w}_{12}(\mathbf{k}_3) = 2w_1 - 6w_3 + \ldots . \qquad (12\text{d})$$

Substitution of expressions (12) into expressions (9a) gives as follows:

$$\lambda_1(\mathbf{0}) = 6w_2 + 2w_4 + |6w_1 + 6w_3| + \ldots,\qquad \lambda_2(\mathbf{0}) = 6w_2 + 2w_4 - |6w_1 + 6w_3| + \ldots; \qquad (13\text{a})$$

$$\lambda_1(\mathbf{k}_1) = -2w_2 + 2w_4 + |-2w_1 + 6w_3| + \ldots,\qquad \lambda_2(\mathbf{k}_1) = -2w_2 + 2w_4 - |-2w_1 + 6w_3| + \ldots; \qquad (13\text{b})$$

$$\lambda_1(\mathbf{k}_2) = -2w_2 + 2w_4 + |2w_1 - 6w_3| + \ldots,\qquad \lambda_2(\mathbf{k}_2) = -2w_2 + 2w_4 - |2w_1 - 6w_3| + \ldots; \qquad (13\text{c})$$

$$\lambda_1(\mathbf{k}_3) = -2w_2 + 2w_4 + |2w_1 - 6w_3| + ..., \quad \lambda_2(\mathbf{k}_3) = -2w_2 + 2w_4 - |2w_1 - 6w_3| + .... \quad (13d)$$

It is easy to see from expressions (13) that

$$\lambda_1(\mathbf{k}_1) = \lambda_1(\mathbf{k}_2) = \lambda_1(\mathbf{k}_3) \equiv \lambda_1(\mathbf{k}^M), \quad \lambda_2(\mathbf{k}_1) = \lambda_2(\mathbf{k}_2) = \lambda_2(\mathbf{k}_3) \equiv \lambda_2(\mathbf{k}^M). \quad (14)$$

As shown in Fig. 2, every Al atom is surrounded only by nearest-neighboring Ti atoms in $D0_{19}$-type superstructure. Thus, the pairwise interaction energy of the nearest Al and Ti atoms is negative. Therefore, as it follows from expression (2), mixing energy is positive for the first co-ordination shell, as for all atomic-ordering systems (see, for instance, [35]), $w_1 > 0$. Assuming that $w_1 > 3w_3$ and using expressions (9) and (13), a correspondence between the above-mentioned eigenvalues and eigenvectors is as follows:

$$\lambda_1(\mathbf{0}) = \tilde{w}_{11}(\mathbf{0}) + \tilde{w}_{12}(\mathbf{0}), \quad \lambda_2(\mathbf{0}) = \tilde{w}_{11}(\mathbf{0}) - \tilde{w}_{12}(\mathbf{0}), \quad \mathbf{v}_1(\mathbf{0}) = \frac{1}{\sqrt{2}}\begin{pmatrix}1\\1\end{pmatrix}, \quad \mathbf{v}_2(\mathbf{0}) = \frac{1}{\sqrt{2}}\begin{pmatrix}1\\-1\end{pmatrix}; \quad (15a)$$

$$\lambda_1(\mathbf{k}_1) = \tilde{w}_{11}(\mathbf{k}_1) - \tilde{w}_{12}(\mathbf{k}_1), \quad \lambda_2(\mathbf{k}_1) = \tilde{w}_{11}(\mathbf{k}_1) + \tilde{w}_{12}(\mathbf{k}_1), \quad \mathbf{v}_1(\mathbf{k}_1) = \frac{1}{\sqrt{2}}\begin{pmatrix}1\\-1\end{pmatrix}, \quad \mathbf{v}_2(\mathbf{k}_1) = \frac{1}{\sqrt{2}}\begin{pmatrix}1\\1\end{pmatrix}; \quad (15b)$$

$$\lambda_1(\mathbf{k}_2) = \tilde{w}_{11}(\mathbf{k}_2) + \tilde{w}_{12}(\mathbf{k}_2), \quad \lambda_2(\mathbf{k}_2) = \tilde{w}_{11}(\mathbf{k}_2) - \tilde{w}_{12}(\mathbf{k}_2), \quad \mathbf{v}_1(\mathbf{k}_2) = \frac{1}{\sqrt{2}}\begin{pmatrix}1\\1\end{pmatrix}, \quad \mathbf{v}_2(\mathbf{k}_2) = \frac{1}{\sqrt{2}}\begin{pmatrix}1\\-1\end{pmatrix}; \quad (15c)$$

$$\lambda_1(\mathbf{k}_3) = \tilde{w}_{11}(\mathbf{k}_3) + \tilde{w}_{12}(\mathbf{k}_3), \quad \lambda_2(\mathbf{k}_3) = \tilde{w}_{11}(\mathbf{k}_3) - \tilde{w}_{12}(\mathbf{k}_3), \quad \mathbf{v}_1(\mathbf{k}_3) = \frac{1}{\sqrt{2}}\begin{pmatrix}1\\1\end{pmatrix}, \quad \mathbf{v}_2(\mathbf{k}_3) = \frac{1}{\sqrt{2}}\begin{pmatrix}1\\-1\end{pmatrix}. \quad (15d)$$

Applying the static concentration waves method, expansion (4) is as follows:

$$\begin{pmatrix}P_1(\mathbf{R})\\P_2(\mathbf{R})\end{pmatrix} = c\begin{pmatrix}1\\1\end{pmatrix} + \eta_{\Gamma 2}\frac{\gamma_{\Gamma 2}(\mathbf{k}^\Gamma)}{\sqrt{2}}\begin{pmatrix}1\\-1\end{pmatrix} +$$

$$+\frac{\eta_{M1}}{2}\left[\frac{\gamma_{M1}(\mathbf{k}_1)}{\sqrt{2}}\begin{pmatrix}1\\-1\end{pmatrix}e^{i\pi\mathbf{a}_1^*\cdot\mathbf{R}} + \frac{\gamma_{M1}^*(\mathbf{k}_1)}{\sqrt{2}}\begin{pmatrix}1\\-1\end{pmatrix}e^{-i\pi\mathbf{a}_1^*\cdot\mathbf{R}}\right] + \frac{\eta_{M2}}{2}\left[\frac{\gamma_{M2}(\mathbf{k}_1)}{\sqrt{2}}\begin{pmatrix}1\\1\end{pmatrix}e^{i\pi\mathbf{a}_1^*\cdot\mathbf{R}} + \frac{\gamma_{M2}^*(\mathbf{k}_1)}{\sqrt{2}}\begin{pmatrix}1\\1\end{pmatrix}e^{-i\pi\mathbf{a}_1^*\cdot\mathbf{R}}\right] +$$

$$+\frac{\eta_{M1}}{2}\left[\frac{\gamma_{M1}(\mathbf{k}_2)}{\sqrt{2}}\begin{pmatrix}1\\1\end{pmatrix}e^{i\pi\mathbf{a}_2^*\cdot\mathbf{R}} + \frac{\gamma_{M1}^*(\mathbf{k}_2)}{\sqrt{2}}\begin{pmatrix}1\\1\end{pmatrix}e^{-i\pi\mathbf{a}_2^*\cdot\mathbf{R}}\right] + \frac{\eta_{M2}}{2}\left[\frac{\gamma_{M2}(\mathbf{k}_2)}{\sqrt{2}}\begin{pmatrix}1\\-1\end{pmatrix}e^{i\pi\mathbf{a}_2^*\cdot\mathbf{R}} + \frac{\gamma_{M2}^*(\mathbf{k}_2)}{\sqrt{2}}\begin{pmatrix}1\\-1\end{pmatrix}e^{-i\pi\mathbf{a}_2^*\cdot\mathbf{R}}\right] +$$

$$+\frac{\eta_{M1}}{2}\left[\frac{\gamma_{M1}(\mathbf{k}_3)}{\sqrt{2}}\begin{pmatrix}1\\1\end{pmatrix}e^{i\pi(\mathbf{a}_1^*+\mathbf{a}_2^*)\cdot\mathbf{R}} + \frac{\gamma_{M1}^*(\mathbf{k}_3)}{\sqrt{2}}\begin{pmatrix}1\\1\end{pmatrix}e^{-i\pi(\mathbf{a}_1^*+\mathbf{a}_2^*)\cdot\mathbf{R}}\right] +$$

$$+\frac{\eta_{M2}}{2}\left[\frac{\gamma_{M2}(\mathbf{k}_3)}{\sqrt{2}}\begin{pmatrix}1\\-1\end{pmatrix}e^{i\pi(\mathbf{a}_1^*+\mathbf{a}_2^*)\cdot\mathbf{R}} + \frac{\gamma_{M2}^*(\mathbf{k}_3)}{\sqrt{2}}\begin{pmatrix}1\\-1\end{pmatrix}e^{-i\pi(\mathbf{a}_1^*+\mathbf{a}_2^*)\cdot\mathbf{R}}\right]. \quad (16)$$

In the completely ordered alloy with a stoichiometric composition, $c = c_{st} = 1/4$, at $T = 0$ K, when the LRO parameters, $\eta_{11}$ and $\eta_{12}$, are equal to 1, function $P_q(\mathbf{R})$ is equal to 0 or 1 over the all sites of the h.c.p. lattice. This condition permits to calculate the symmetry coefficients, $\gamma_{\Gamma 2}(\mathbf{k}_0)$, $\gamma_{M1}(\mathbf{k}_1)$, $\gamma_{M1}^*(\mathbf{k}_1)$, $\gamma_{M2}(\mathbf{k}_1)$, $\gamma_{M2}^*(\mathbf{k}_1)$, $\gamma_{M1}(\mathbf{k}_2)$, $\gamma_{M1}^*(\mathbf{k}_2)$, $\gamma_{M2}(\mathbf{k}_2)$, $\gamma_{M2}^*(\mathbf{k}_2)$, $\gamma_{M1}(\mathbf{k}_3)$, $\gamma_{M1}^*(\mathbf{k}_3)$, $\gamma_{M2}(\mathbf{k}_3)$, $\gamma_{M2}^*(\mathbf{k}_3)$, for all types of the crystalline structure based on h.c.p. lattice with a stoichiometry ($c = 1/4$).

Taking into account properties of the basic vectors, the single-site (occupation-probability) function of distribution, $P_q(\mathbf{R})$, for the $D0_{19}$-type superstructure becomes as follows:

$$\begin{pmatrix} P_1(\mathbf{R}) \\ P_2(\mathbf{R}) \end{pmatrix} = c + \eta \begin{pmatrix} E_1(\mathbf{R}) \\ E_2(\mathbf{R}) \end{pmatrix}, \text{ where } \begin{pmatrix} E_1(\mathbf{R}) \\ E_2(\mathbf{R}) \end{pmatrix} = \frac{1}{4}\left[\xi_1\begin{pmatrix} 1 \\ 1 \end{pmatrix}e^{i\pi\mathbf{a}_1^*\cdot\mathbf{R}} + \xi_2\begin{pmatrix} 1 \\ -1 \end{pmatrix}e^{i\pi\mathbf{a}_2^*\cdot\mathbf{R}} + \xi_3\begin{pmatrix} 1 \\ -1 \end{pmatrix}e^{i\pi(\mathbf{a}_1^*+\mathbf{a}_2^*)\cdot\mathbf{R}}\right]; \quad (17)$$

where $\xi_1 = \xi_2 = \xi_3 = 1$, $\xi_1 = -\xi_2 = -\xi_3 = 1$, $-\xi_1 = \xi_2 = -\xi_3 = 1$, or $-\xi_1 = -\xi_2 = \xi_3 = 1$. Function (17) is dependent on one LRO parameter, $\eta$, and assumes two values, $c - \eta/4$ and $c + 3\eta/4$, on all crystal lattice sites, i.e. the function satisfies the corresponding criterion formulated by Khachaturyan [26].

Substitution expression (17) into expression (1) yields the configurational free energy (per atom) of ordered $D0_{19}$ phase as a function of temperature, concentration and the nonzero LRO parameter:

$$F_{ord} = c^2\lambda_1(\mathbf{0})/2 + 3\eta^2\lambda_2(\mathbf{k}^M)/32 + (k_BT/4)[(c+3\eta/4)\ln(c+3\eta/4) +$$

$$+(1-c-3\eta/4)\ln(1-c-3\eta/4) + 3(c-\eta/4)\ln(c-\eta/4) + 3(1-c+\eta/4)\ln(1-c+\eta/4)]. \quad (18)$$

The configurational free energy (per atom) of the disordered phase (where $\eta \equiv 0$) is

$$F_{disord} = c^2\lambda_1(\mathbf{0})/2 + k_BT[c\ln c + (1-c)\ln(1-c)]. \quad (19)$$

The equilibrium fields of the ordered ($\alpha_2$) and disordered ($\alpha$) phases, which are the parts of a phase diagram of the Ti–Al system, are determined by the interaction-energy parameters, $\lambda_1(\mathbf{0})$ and $\lambda_2(\mathbf{k}^M)$, which enter into the free energy expressions, (18) and (19). The energy values can be obtained in some ways: from the radiation (x-rays or thermal neutrons) scattering data (however until now, these measurements apparently have not been done for Ti–Al alloys), from the first-principles' calculations, or by fitting interaction parameters to the available experimental phase diagram [36-38]. It was chosen the last way – the fitting procedure.

**Kinetics Model of Long-Range Order for $D0_{19}$ Phase**

Let us consider now the case of exchange ('ring') mechanism [27,29–35] governing the diffusion relaxation of binary substitutional h.c.p. solid solution under its ordering.

In $D0_{19}$(Ti$_3$Al)-type phase, predominance of this diffusion mechanism is conditioned by a very low vacancy concentration [13]. For instance, the vacancy amount referred to the total number of lattice sites is found to be about 0.0009 in the exactly stoichiometric Ti$_3$Al phase at 850°C [13]. This means that the degree of atomic order as well as the presence of antistructure atoms are conditioned almost entirely by the deviation from stoichiometry and processing environment.

To investigate the kinetics of an atomic-ordering process in (non-equilibrium) h.c.p.-$A_{1-c}B_c$ solid solution, a model based on the Onsager-type microscopic-diffusion equation [26] was taken. The rates of change of single-site probabilities for $B$-atoms (within the first and second co-ordination sublattices) are as follows:

$$\frac{dP_1(\mathbf{R},t)}{dt} = -\frac{1}{k_BT}\sum_{\mathbf{R}'}\left\{\left[c^2 L_{11}^{BB}(\mathbf{R}-\mathbf{R}') - c(1-c)L_{11}^{BA}(\mathbf{R}-\mathbf{R}')\right]\frac{\delta F}{\delta P_1(\mathbf{R}',t)} +\right.$$

$$\left. + \left[c^2 L_{12}^{BB}(\mathbf{R}-\mathbf{R}') - c(1-c)L_{12}^{BA}(\mathbf{R}-\mathbf{R}')\right]\frac{\delta F}{\delta P_2(\mathbf{R}',t)}\right\}, \quad (20a)$$

$$\frac{dP_2(\mathbf{R},t)}{dt} = -\frac{1}{k_B T}\sum_{\mathbf{R}'}\left\{\left[c^2 L_{11}^{BB}(\mathbf{R}-\mathbf{R}') - c(1-c)L_{11}^{BA}(\mathbf{R}-\mathbf{R}')\right]\frac{\delta F}{\delta P_2(\mathbf{R}',t)} + \right.$$
$$\left. +\left[c^2 L_{12}^{BB}(\mathbf{R}-\mathbf{R}') - c(1-c)L_{12}^{BA}(\mathbf{R}-\mathbf{R}')\right]\frac{\delta F}{\delta P_1(\mathbf{R}',t)}\right\}. \quad (20b)$$

Here, $t$ is a time, $L_{pq}^{\alpha\beta}(\mathbf{R}-\mathbf{R}')$ is a matrix of the kinetic coefficients whose elements represent exchange probabilities for an elementary diffusion jumps of a pair of atoms, $\alpha$ and $\beta$, between the site $\mathbf{r}$ of the $p$-th sublattice and site $\mathbf{r}'$ of the $q$-th sublattice during the time unit ($\alpha, \beta = A, B$; $p, q = 1, 2$). (Since such a probability for a pair of sites at $\mathbf{r} = \mathbf{R} + \mathbf{h}_p$ and $\mathbf{r}' = \mathbf{R}' + \mathbf{h}_q$ is invariant under Bravais translations only, its dependence on difference of Bravais-translation vectors, $(\mathbf{R}-\mathbf{R}')$ is obtained) Note also that for exchange diffusion mechanism it is enough to consider migration, for example, of only $B$-atoms, since a sum of occupation probabilities for $A$ and $B$ atoms is identically equal to unity. The Onsager reciprocity relations, $L_{11}^{BB}(\mathbf{R}-\mathbf{R}') = L_{22}^{BB}(\mathbf{R}-\mathbf{R}')$, $L_{11}^{BA}(\mathbf{R}-\mathbf{R}') = L_{22}^{BA}(\mathbf{R}-\mathbf{R}')$ $L_{12}^{BB}(\mathbf{R}-\mathbf{R}') = L_{21}^{BB}(\mathbf{R}-\mathbf{R}')$, $L_{12}^{BA}(\mathbf{R}-\mathbf{R}') = L_{21}^{BA}(\mathbf{R}-\mathbf{R}')$ are postulated (for $D0_{19}$-type phase nearby the equilibrium) and taken into account in Eqns. 20a and 20b.

Since the total number of $B$ ($A$) atoms is fixed, there is a following restriction for the kinetic coefficients:

$$\sum_{p=1}^{2}\sum_{\mathbf{R}}\frac{dP_p(\mathbf{R})}{dt} = \frac{dN_B}{dt} \equiv 0, \text{ i.e. } \sum_{p=1}^{2}\sum_{\mathbf{R}}L_{pq}^{\alpha\beta}(\mathbf{R}-\mathbf{R}') = 0, \quad (21)$$

where $N_B$ is the total number of $B$ atoms in a system.

The thermodynamic driving forces, $\delta F/\delta P_q(\mathbf{R}',t)$, can be written as:

$$\frac{\delta F}{\delta P_1(\mathbf{R}')} = \sum_{\mathbf{R}}[w_{11}(\mathbf{R}-\mathbf{R}')P_1(\mathbf{R}) + w_{12}(\mathbf{R}-\mathbf{R}')P_2(\mathbf{R})] + k_B T \ln\frac{P_1(\mathbf{R}')}{1-P_1(\mathbf{R}')}, \quad (22a)$$

$$\frac{\delta F}{\delta P_2(\mathbf{R}')} = \sum_{\mathbf{R}}[w_{11}(\mathbf{R}-\mathbf{R}')P_2(\mathbf{R}) + w_{12}(\mathbf{R}-\mathbf{R}')P_1(\mathbf{R})] + k_B T \ln\frac{P_2(\mathbf{R}')}{1-P_2(\mathbf{R}')}. \quad (22b)$$

The $\delta F/\delta P_q(\mathbf{R})$ value possesses the same symmetry as the function $P_q(\mathbf{R})$; therefore in case of the LRO, the $\delta F/\delta P_q(\mathbf{R})$ value as well as the $P_q(\mathbf{R})$ value can be represented as the superpositions of the same concentration waves [26]:

$$\delta F/\delta P_q(\mathbf{R}) = \tilde{c}(\eta) + \tilde{\eta}(\eta)E_q(\mathbf{R}), \quad P_q(\mathbf{R}) = c + \eta E_q(\mathbf{R}) \quad (q = 1, 2); \quad (23)$$

here, $E_q(\mathbf{R})$ are given by expression 17 (for $D0_{19}$ structure). Combining expressions (22), (23) and taking into account that $E_q(\mathbf{R})$ takes only two values, $-1/4$ and $3/4$, at the all h.c.p. lattice sites, it is easy to see that the functions $\tilde{c}(\eta)$ and $\tilde{\eta}(\eta)$ are as follows:

$$\tilde{c}(\eta) = c\lambda_1(\mathbf{0}) + \frac{k_B T}{4}\ln\frac{(c-\eta/4)^3(c+3\eta/4)}{(1-c+\eta/4)^3(1-c-3\eta/4)}, \quad (24a)$$

$$\tilde{\eta}(\eta) = \eta\lambda_2(\mathbf{k}^M) + k_B T \ln\frac{(c+3\eta/4)(1-c+\eta/4)}{(1-c-3\eta/4)(c-\eta/4)}. \quad (24b)$$

Substituting expressions (23) into Eqs. 20, combining equalities (21) and (24), followed by the Fourier transformation of both members of Eqs. 20, the differential equation for the true LRO parameter, η is obtained:

$$\frac{d\eta}{dt} = -c(1-c)\tilde{L}(\mathbf{k}^M)\left[\frac{\lambda_2(\mathbf{k}^M)}{k_B T}\eta + \ln\frac{(c+3\eta/4)(1-c+\eta/4)}{(1-c-3\eta/4)(c-\eta/4)}\right], \quad (25)$$

where

$$\tilde{L}(\mathbf{k}^M) = \frac{c}{1-c}\left[\tilde{L}_{11}^{BB}(\mathbf{k}^M) + \tilde{L}_{12}^{BB}(\mathbf{k}^M)\right] - \left[\tilde{L}_{11}^{BA}(\mathbf{k}^M) + \tilde{L}_{12}^{BA}(\mathbf{k}^M)\right]. \quad (26)$$

Thus, at least if $c \ll 1$, an exchange of the same-kind atoms does not change η. As easy to see from Eq. 25, it is conveniently to define both a reduced time $t^* = \tilde{L}(\mathbf{k}^M)t$ and a reduced temperature $T^* = k_B T/|\lambda_2(\mathbf{k}^M)|$ to solve the kinetics equation numerically in terms of $t^*$ and $T^*$.

**Results and Conclusions**

**Statistical-Thermodynamic Calculations.** According to the experimental phase diagram [36–38] in Fig. 3, there is a concentration–temperature range, where the equilibrium state of Ti–Al system corresponds to a disordered solid state (α-phase). With increasing the Al concentration or/and decreasing the temperature, this disordered phase becomes unstable, and the ordered α$_2$-phase appears.

To calculate the equilibrium LRO parameter, the configurational free energy, F, with respect to η has to be minimized. For $c = 1/4$, such a procedure performed at different temperatures gives the A–B–E branch in Fig. 4a. On the other hand, the equilibrium LRO parameter, η$_{eq}$, must satisfy the following condition for F being minimum with respect to η:

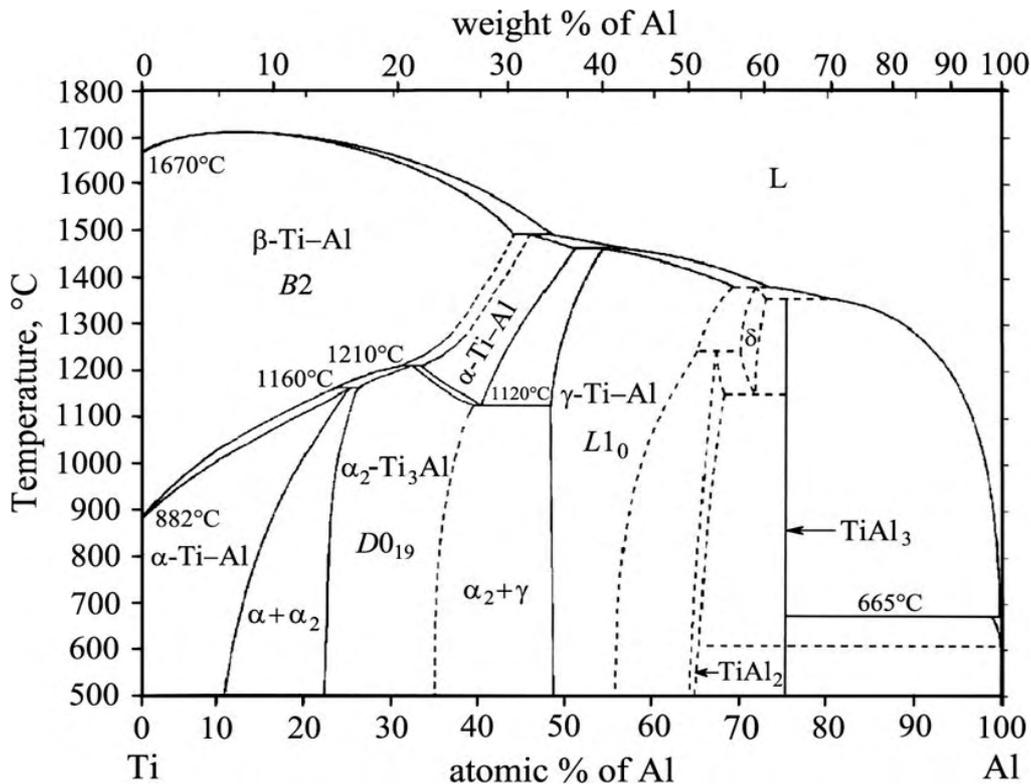

Fig. 3 Phase diagram for Ti–Al alloy [36–38].

$$\frac{\eta_{eq}}{T^*} = \ln \frac{(c - \eta_{eq}/4)(1 - c - 3\eta_{eq}/4)}{(c + 3\eta_{eq}/4)(1 - c + \eta_{eq}/4)}. \qquad (27)$$

The last equation has no simple analytic solution, but one can solve it numerically. The solution, for a given $c = 1/4$, has a form of the $A$–$B$–$F$ branch shown in Fig. 4a.

The behavior of the equilibrium LRO parameter (Fig. 4a) is discussed by dividing the reduced-temperature axis, $T^*$, into several ranges, which are separated by the temperatures $T^*_F$, $T^*_E$ and $T^*_G$ (corresponding points are indicated in Fig. 4a). When $T^* > T^*_G$, $\eta_{eq} \equiv 0$, and the disordered solution is stable. When $T^*_G > T^* > T^*_E$, Eq. 27 has three solutions, $\eta_{eq} \equiv 0$, $\eta_{eqC-D} \neq 0$ and $\eta_{eqC-B} \neq 0$ (see Fig. 4a). Within this temperature interval, there is a coexistence of both ordered phase ($\eta_{eqC-B} \neq 0$) and disordered one ($\eta_{eq} \equiv 0$). However, the free energy of disordered phase is lower, and this phase is more stable. Within this temperature domain, the ordered structure with $\eta_{eqC-B} \neq 0$ is metastable, and the states with $\eta_{eqC-D} \neq 0$ are labile. When $T^*_E > T^* > T^*_F$, the last equation has the same three solutions, nevertheless situation is inverse: there is a coexistence of two phases as well, but now $\eta_{eq} = \eta_{eqB-A}$ only (not $\eta_{eq} = \eta_{eqD-F}$) provides the least values of free energy, and ordered phase along the B–A branch is more stable. If $T^* < T^*_F$, there are two solutions: $\eta_{eq} \equiv 0$ and $\eta_{eq} \neq 0$; this solution $\eta_{eq} \neq 0$ minimizes the free energy while $\eta_{eq} \equiv 0$ maximizes it. Thus, the ordered phase is stable, and disordered solution is absolutely unstable.

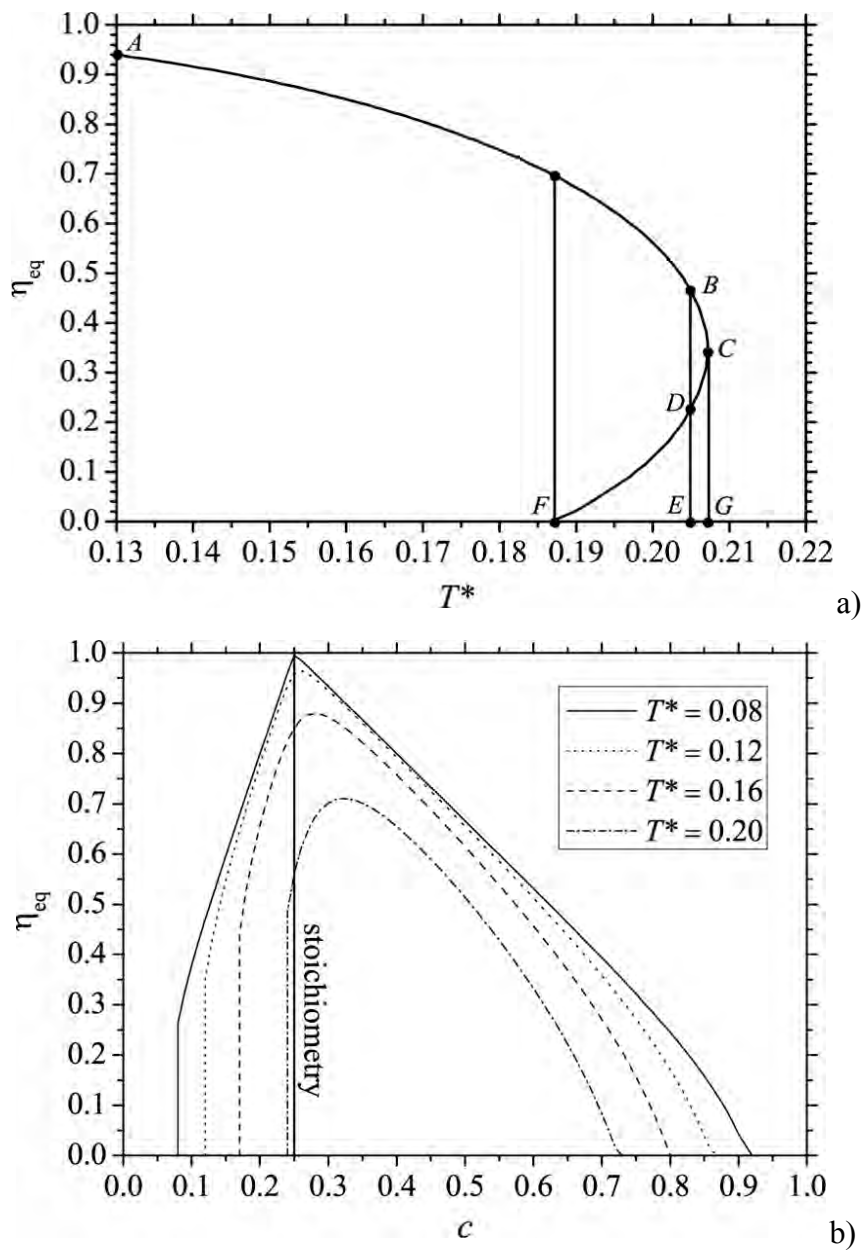

Fig. 4 Equilibrium LRO parameter vs. the reduced temperature (a) and vs. the concentration (b) for $D0_{19}$-type phase, which is stoichiometric ($c = 1/4$) in the left figure.

Figure 4a shows that phase transformation of disordered h.c.p. solid solution into the ordered $D0_{19}$-phase is the first-kind phase transition (as, e.g., $\alpha \to \alpha_2$ phase transition in h.c.p.-Ti–Al).

The equation $\partial F/\partial \eta = 0$ also yields the concentration dependence of the equilibrium LRO parameter, $\eta_{eq} = \eta_{eq}(c)$. The curves in Fig. 4b represent the numerical solutions of this equation for the different reduced temperatures, $T^*$.

If the equilibrium LRO parameter is known, a configurational free energy can be calculated as a function of temperature and concentration. The equilibrium compositions for the coexistent disordered $\alpha$-Ti–Al phase and ordered intermetallic $\alpha_2$-Ti$_3$Al-type phase (see Fig. 3) can be determined numerically by the common tangent construction. In this case, the values of $\lambda_1(\mathbf{0})$ and $\lambda_2(\mathbf{k}^M)$ in Eqns. 18 and 19 are the fitting parameters, which was estimated for arbitrary interatomic distances (i.e. outside the scope of the conventional Bragg–Williams approximation). Using this procedure combined with the least-squares method, the phase relationships were computed for two assumptions (Fig. 5a). The first approximation yields the temperature-independent eigenvalues of the interchange-energy matrix ($\lambda_1(\mathbf{0}) \cong 462.36$ meV/atom, $\lambda_2(\mathbf{k}^M) \cong -555.91$ meV/atom) relevant to the whole temperature range of 600–1100°C. The second one gives temperature-dependent values ($\lambda_1(\mathbf{0}) = \lambda_1(\mathbf{0},T)$, $\lambda_2(\mathbf{k}^M) = \lambda_2(\mathbf{k}^M,T)$). The fitted temperature dependences of both $\lambda_1(\mathbf{0},T)$ and $\lambda_2(\mathbf{k}^M,T)$ are shown in Fig. 5b. The calculations were not extended for more high temperatures, since, at higher temperatures, the Ti$_3$Al-type phase coexists already with the b.c.c. β-phase.

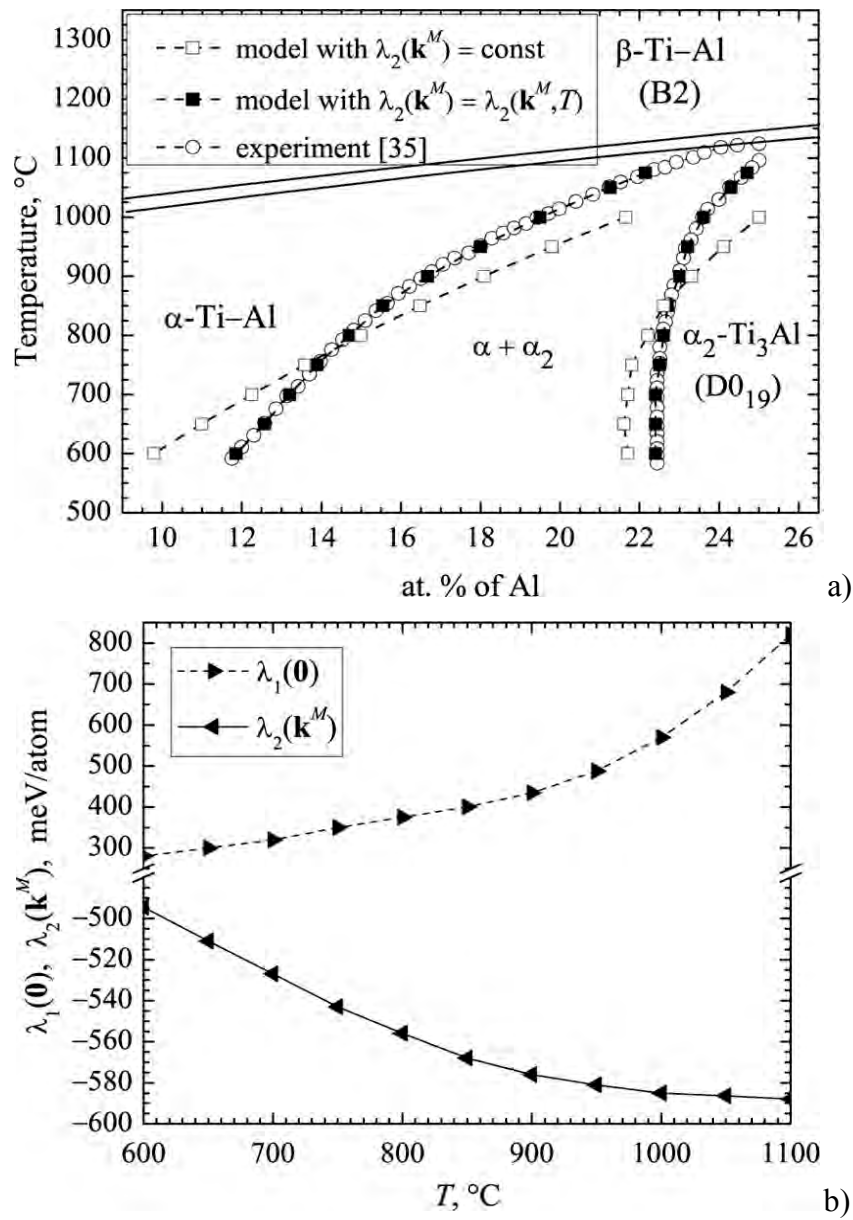

Fig. 5 Calculated and experimental partial phase diagrams (a) and temperature dependence of interatomic-interaction parameters for h.c.p. Ti–Al alloy.

A change of the lattice parameters, $a_0$ and $c_0$, with both temperature and composition results in the change of interatomic-interaction energies, which are implicitly dependent on $T$ and $c$. In particular, the thermal expansion, the temperature dependence of elasticity, and the softening of oscillatory modes are responsible for the temperature dependence of both $\lambda_1(\mathbf{0})$ and $\lambda_2(\mathbf{k}^M)$.

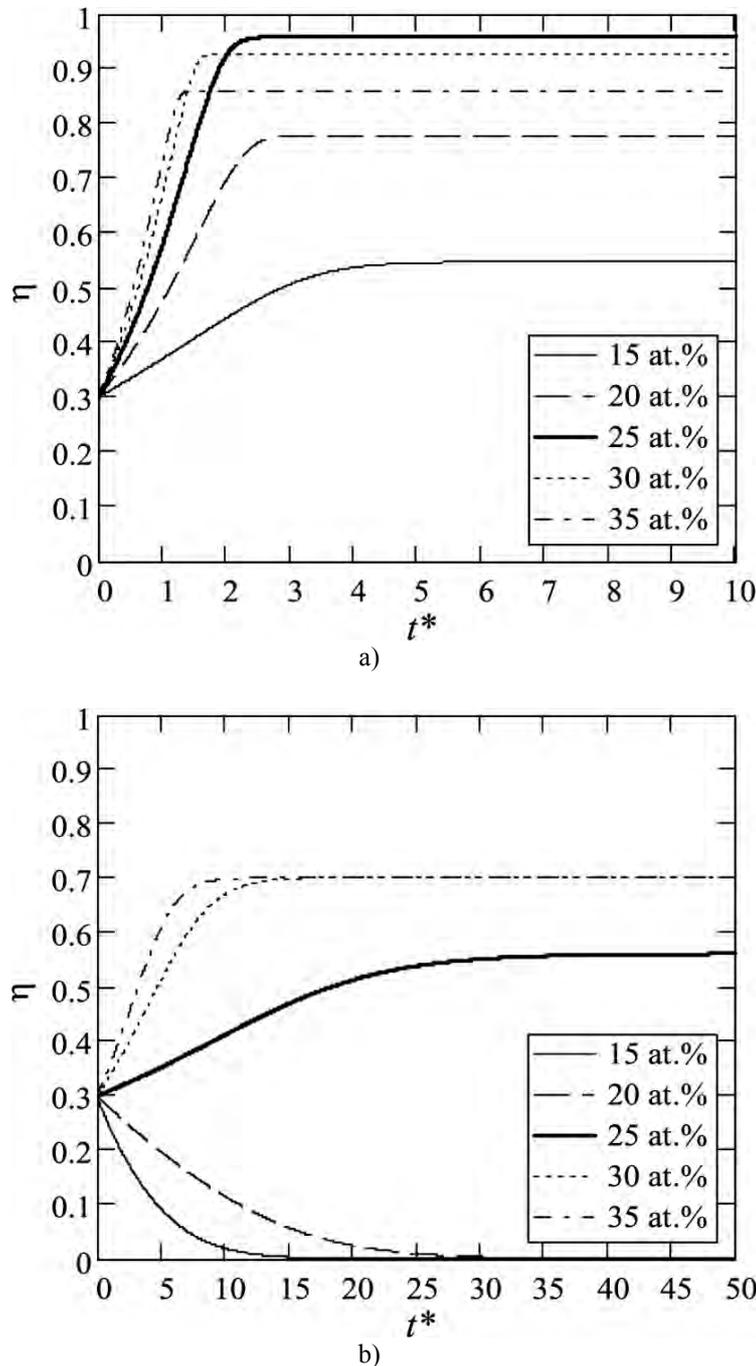

Fig. 6 The reduced-time dependences of the LRO parameter for different atomic percents of alloying component in D0$_{19}$-type $A_{1-c}B_c$ alloy at the reduced temperatures: $T^* = 0.12$ (a), $T^* = 0.20$ (b).

According to Fig. 5a, for the model with temperature-dependent interatomic-interaction parameters, the computed phase boundaries almost coincide with the experimental findings; and for the model with temperature-independent interatomic-interaction parameters, there is a less agreement.

**Kinetic Calculations.** In Figures 6a, 6b, curves represent the numerical solutions of Eq. 25 at initial magnitude of the LRO parameter, $\eta_0 = \eta(t=0) = 0.3$, for different atomic fractions of alloying component ($B$).

Figures 6a and 6b demonstrate how both the concentration and the annealing temperature affect the quantitative and qualitative changes of the kinetic and equilibrium paths. Rise in temperature results in both the decelerating of the LRO parameter change (at the initial evolution stage) and the increasing of a relaxation time. As shown in Fig. 6b, $\eta_{eq} = 0$ for $A_{0.85}B_{0.15}$ and $A_{0.80}B_{0.20}$ alloys at $T^* = 0.20$. It means that disordering occurs at such 'temperatures'. Equilibrium values of the LRO parameter in Figs. 6a, 6b coincide with corresponding values of $\eta_{eq}$ in Fig. 4b. Thus, both the kinetics and statistical-thermodynamic models give equal values of the equilibrium LRO parameter, i.e. they unambiguously define a state (ordered or disordered one) the system is in.

At low temperatures (for instance, at $T^* = 0.12$ in Fig. 6a), an equilibrium LRO parameter for non-stoichiometric D0$_{19}$-type $A_{1-c}B_c$ alloys with both $c < 1/4$ and $c > 1/4$ is always lower than for stoichiometric one ($c = 1/4$). However, an equilibrium LRO parameter for non-stoichiometric alloys with $c > 1/4$ can be higher than for a stoichiometric one at high $T$ (for instance, at $T^* = 0.20$ in Fig. 6b). This is confirmed by the statistical-thermodynamic results in Fig. 4b.

Assuming the temperature independence of interatomic-interaction parameter, $\lambda_2(\mathbf{k}^M) = -555.91$ meV/atom, in Eq. 25, a relaxation kinetics of the LRO parameter was calculated for stoichiometric Ti$_3$Al alloy ($c = 1/4$). Figure 7a shows results of these calculations for $\eta_0 = 0.25$ (when $t=0$). For comparison, Fig. 7b shows a relaxation kinetics of the LRO parameter if $\lambda_2(\mathbf{k}^M)$ has a temperature dependence as in Fig. 5b. According to Figs. 7a and 7b, the energy factor has a significant influence on the relaxation kinetics. The two types of calculations give not only different profiles of the relaxation curves but also the equilibrium LRO-parameter values. Our results demonstrate that the interatomic-interaction parameters strongly affect quantitative changes as well as qualitative ones of the kinetic and equilibrium paths. Comparison between Figs. 7a and 7b demonstrates that curves in Fig. 7b draw together along the vertical line. It means that a temperature influence on the interatomic interactions diminishes the spread in the values of instantaneous and equilibrium LRO parameters versus the temperature.

Using $T$-dependent $\lambda_2(\mathbf{k}^M)$, Eq. 25 is solved for different concentrations of Al in Ti–Al alloy. In Figs. 8a and 8b, curves represent the solutions. The concentrations, $c$, are chosen only for those regions of the Ti–Al phase diagram [36–38] in Fig. 3, which have an h.c.p. structure as a whole, i.e. for the two-phase region, $\alpha + \alpha_2$, and for the single-phase one, $\alpha_2$. The two-phase region $\alpha_2 + \gamma$ is not considered because the face centred cubic structure ($\gamma$-phase) is out of the consideration in a given paper.

It is easy to see in Figs. 8a and 8b that an initial magnitude of the LRO parameter is by no means influencing on the equilibrium LRO parameter—its magnitude is the same for all $\eta_0$ at a given $T$.

As shown in Fig. 8, $\eta_{eq} = 0$ for Ti$_{0.85}$Al$_{0.15}$ at 900ºC. The experimental phase diagrams [36–38] confirm this theoretical conclusion; in the phase diagram in Fig. 3, corresponding point belongs to the disordered region. To the contrary, the equilibrium atomic arrangements in Ti$_{0.80}$Al$_{0.20}$, Ti$_{0.75}$Al$_{0.25}$ and Ti$_{0.70}$Al$_{0.30}$ alloys at both 500ºC and 900ºC are ordered (Fig. 8) that is in accordance with experimental phase diagrams [36–38].

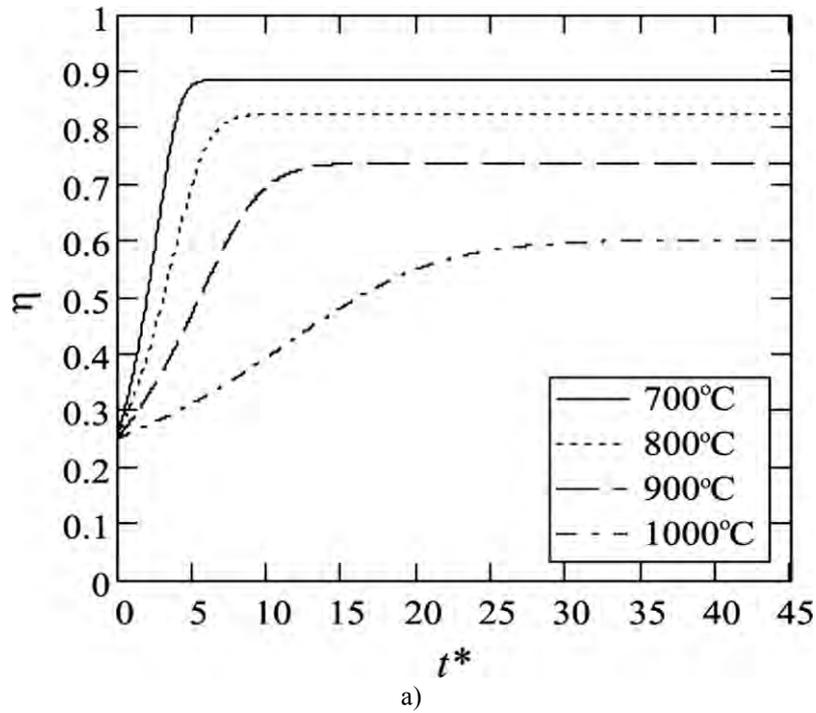

a)

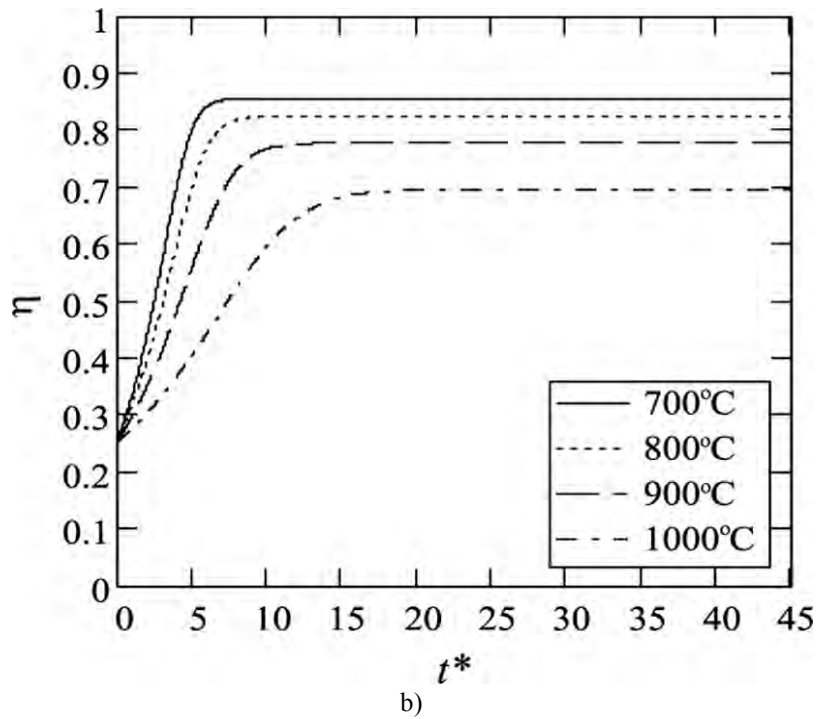

b)

Fig. 7 Reduced-time dependence of the LRO parameter for a stoichiometric Ti$_3$Al phase at the different temperatures within the scope of the model with temperature-independent interatomic-interaction parameters (a) and with temperature-dependent ones (b).

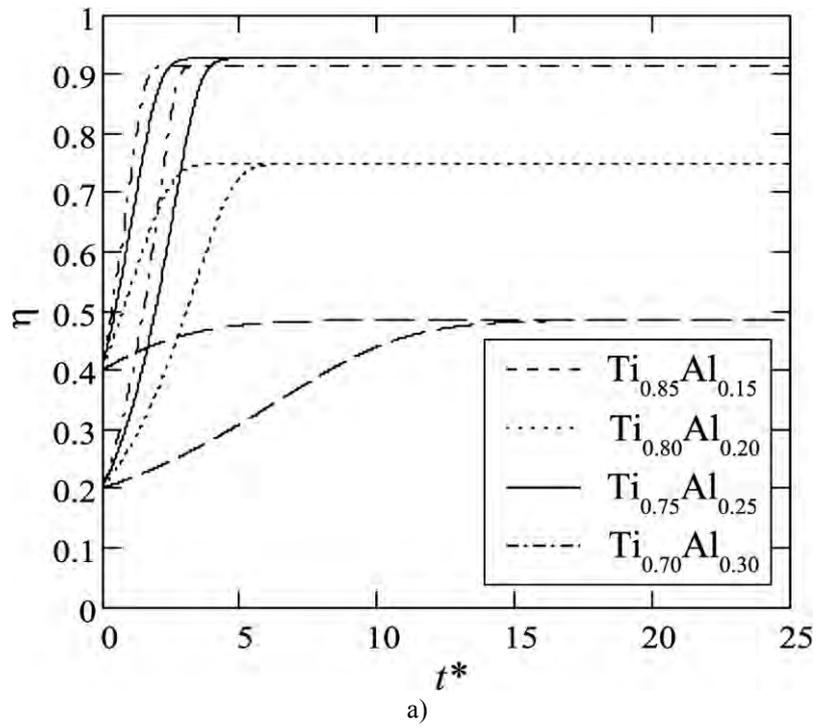

a)

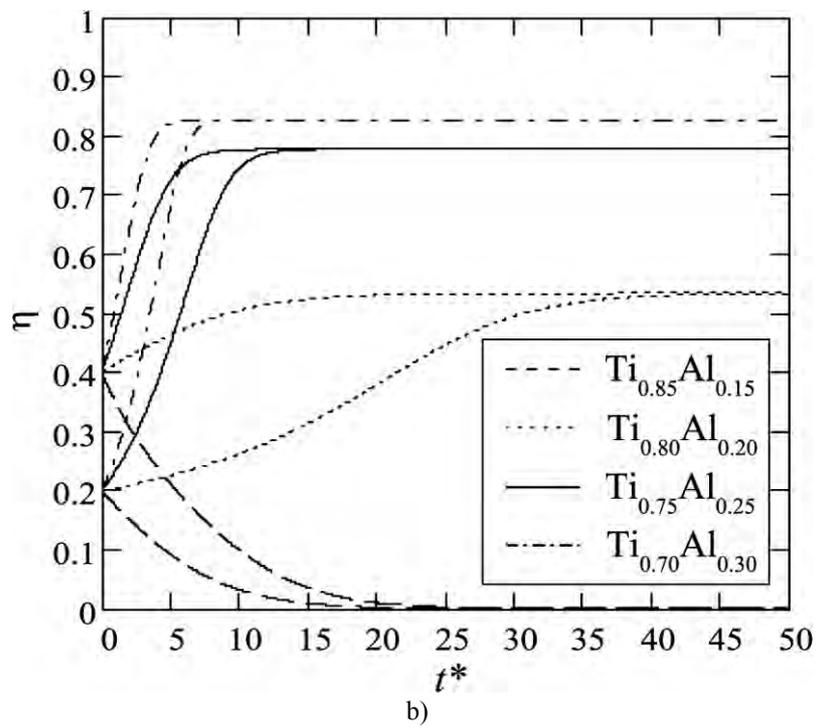

b)

Fig. 8 The LRO parameter vs. the reduced time for h.c.p. $Ti_{1-c}Al_c$ at 500ºC (a) and 900ºC (b).

**Summary**

The study deals with a semi-phenomenological description of the disorder–order phase transformation of the h.c.p. solid solution into the ordered $D0_{19}$-type phase using both the self-consistent field approximation and the method of static concentration waves. The atomic configurations of the ordered state are described by single-site occupation probability functions, which have been derived for the $D0_{19}$-type structure. The findings eliminate the disagreement in the literature as regards the atomic-distribution functions for the $D0_{19}$ superstructure. As a case in point, the proposed statistical-thermodynamics and kinetics models were applied for h.c.p. Ti–Al alloy.

By computing the partial equilibrium Ti–Al phase diagram (Fig. 5a), interatomic-interaction parameters were evaluated. For the model with temperature-dependent interatomic-interaction parameters, the computed phase boundaries almost coincide with the experimental findings. For the model with temperature-independent interaction parameters, there is a less agreement (Figs. 5a, 5b).

The time-evolution equation for the $D0_{19}$-type LRO parameter in a binary h.c.p. phase is derived on the basis of the Onsager-type microscopic-diffusion equation. It permitted to obtain the time dependence of the LRO parameter for a wide temperature–concentration range.

The physical-kinetics results (Figs. 6a, 6b) confirm the statistical-thermodynamics ones (Fig. 4b). Firstly, the equilibrium LRO parameters coincide within the both models. Secondly, for non-stoichiometric alloys (where an atomic fraction of alloying component is more than 25%), $\eta_{eq}$ can be higher than it is for stoichiometric ones at high temperatures.

The temperature influence on the interatomic interactions (Fig. 5b) changes the relaxation time, namely accelerates the LRO relaxation and diminishes the spread in the values of instantaneous and equilibrium LRO parameters versus the temperature (Figs. 7a, 7b).

As expected, an initial magnitude of the LRO parameter governs a relaxation time of the LRO parameter, but does not determine its equilibrium value (Figs. 8a, 8b).

The LRO parameter relaxes to its equilibrium value, $\eta_{eq}(T,c)$, predicting the state (ordered or disordered) the phase is in. The experimental phase diagrams [36–38] (Fig. 3) confirm the predicted states for h.c.p. Ti–Al (Fig. 8).

**Acknowledgements**

The work was performed within the framework of the project supported by the NATO Reintegration Grant (RIG 981326), which is gratefully acknowledged by the first author. The research used resources of the Computer Centre CRIHAN of Haut-Normandie. Authors are thankful to Prof. A.G. Khachaturyan for very useful discussions of this paper.